\documentclass[aps,prx,twocolumn,groupedaddress,floats,showpacs,final]{revtex4-1}

\usepackage{graphicx}
\usepackage[colorlinks]{hyperref}
\usepackage{makecell}
\usepackage{mathrsfs}
\usepackage{dcolumn}
\usepackage{amsmath}

\begin{document}


\title{Quantized Magnetic Flux and the Magneto-halon Effect in a Critical Superconductor}


\author{Kun Chen}
\affiliation{Department of Physics and Astronomy, Rutgers, The State University of New Jersey Piscataway, NJ 08854-8019, USA}

\author {Boris Svistunov}
\affiliation{Department of Physics, University of Massachusetts, Amherst, Massachusetts 01003, USA}
\affiliation{National Research Center ``Kurchatov Institute," 123182 Moscow, Russia }
\affiliation{Wilczek Quantum Center, School of Physics and Astronomy and T. D. Lee Institute, Shanghai Jiao Tong University, Shanghai 200240, China}
\date{\today}

\date{\today}

\begin{abstract}
Employing the standard worldline-vortex mapping, we conclude that at the critical temperature, superconductors demonstrate
the magneto-halon effect with respect to the quantized net magnetic flux generated by a solenoid inserted into the system. The effect is a direct counterpart of the recently revealed halon effect in terms of the quantized particle charge of a static impurity in the two-dimensional U(1) quantum-critical environment.
The flux-loop model (a.k.a. frozen lattice superconductor) with a quasi-solenoid perturbation proves to be the model of choice for qualitative 
and quantitative description of both the halon and the magneto-halon effects.
\end{abstract}


\maketitle


 In its standard setup, the phenomenon of magnetic flux quantization implies that one is dealing with a persistent
current in a massive toroidal superconductor; with the theoretical proof of the quantization based on the effect of expulsion 
of the magnetic field---along with the electric current---from the bulk of the system (see, e.g., text~\cite{book}).
At the critical temperature, the magnetic penetration length diverges thus rendering the proof inapplicable; furthermore, the very setup becomes meaningless in view of the absence of persistent-current states. One might expect then that at the critical temperature, there is simply no context for the effect of magnetic flux quantization. However, such an expectation proves wrong. 

At the critical temperature, there is still a legitimate setup involving a solenoid inserted into the system and naturally raising the question
of whether the net magnetic flux is still quantized. If the quantization of magnetic flux does take place, an even more intriguing question emerges:
What is the scenario of the transition---controlled by the bare flux of the solenoid---between two adjacent values of quantized magnetic flux?

The answer to the question readily follows by the well-known exact mapping \cite{Halperin1,Halperin2}
(for a textbook discussion, see Ref.~\cite{book}) between the vortex lines of a three-dimensional superconductor and the worldlines of $(2+1)$-dimensional 
neutral bosons in the ground state. The mapping reduces the problem at hand to the (recently discussed) problem of the particle charge of a static impurity 
in the two-dimensional U(1) quantum-critical bosonic environment, where the halon effect takes place \cite{kun1,Whitsitt_Sachdev,kun2}. 

The halon is a special critical state of an impurity in a quantum-critical environment. The hallmark of the halon physics is  the fractionalization of a well-defined integer charge into two parts: a microscopic core with half-integer charge and a critically large halo carrying a complementary charge of $\pm 1/2$. The halon phenomenon emerges when the impurity-environment interaction is fine-tuned to the vicinity of a boundary quantum critical point (BQCP), at which the energies of two quasiparticle states with adjacent integer charges approach each other.
The universality class of such BQCP is captured by a model of pseudo-spin-$1/2$ impurity coupled to the quantum-critical environment, in such a way that the rotational symmetry in the pseudo-spin $xy$-plane is respected, with a small local ``magnetic" field along the pseudo-spin $z$-axis playing the role of control parameter driving the system away from the BQCP. On the approach to BQCP, the half-integer projection of the pseudo-spin on its $z$-axis gets delocalized into a halo of critically divergent radius, capturing the essence of the phenomenon of charge fractionalization.

In terms of the worldline-vortex mapping, the critical point of the finite-temperature superconducting transition in 3D corresponds to the point 
of superfluid--Mott-insulator quantum phase transition in 2D.  All by itself, the mapping implies the existence of a finite-temperature counterpart of the
halon effect: A certain straight-line ``impurity" in a critical superconductor---a counterpart of a static impurity in the bosonic quantum-critical environment---should be able to induce a boundary phase transition controlled by the strength of the coupling to the environment. 

A more subtle question is that of the counterpart of the halo charge, as well as the related question of the physical interpretation of the line impurity. 
We argue that the counterpart of the particle charge is nothing but the quantized magnetic flux. 
Correspondingly, the simplest realization of the line impurity is an infinitesimally thin solenoind.
The subtlety comes from the following circumstance. With bosons, the notion of the charge quantum---a particle---is fundamentally microscopic. 
In particular, it is hardwired  in the worldline representation, where a worldline is associated with a particle (or a hole), no matter what is the state of the system.
In a superconductor, the counterpart of the particle worldline is the vortex line. The property (of an infinitely large vortex line) to carry quantized magnetic flux is {\it emergent} and is not supposed to apply to small vortex loops. 
The property follows from the effect of complete suppression of the circulation of the current away from the vortex line. 
The effect is readily traced
in the case of a single vortex line in a superconducting phase, but is not {\it a priori} obvious in the normal phase, or at the critical point.  
Nevertheless, based on the worldline-vortex mapping, one can argue that an infinitely large vortex line carries exactly one quantum of magnetic flux, no matter what is the state of the system. Indeed, according to the mapping, the vortex-vortex interaction in the charged U(1) matter field is essentially short-ranged. 
In the absence of complete suppression of the circulation of the velocity away from the vortex line, that would be impossible, since finite circulation of the velocity field is known to be responsible for long-range interaction between elements of vortex lines. Last but not least, note that microscopic vortex loops---while having nothing to do with quantized magnetic flux---cannot change the expectation value of the net magnetic flux associated with the solenoid (or any other line impurity). This is an immediate consequence of the combination of the zero divergence and an essentially local character of the magnetic field produced by a microscopic vortex loop.

The microscopic details of the impurity in the 2D quantum-critical environment---the counterpart of the solenoid in the 3D critical superconductor---are not important, provided the global U(1) symmetry of the system is respected. Especially convenient---in a number of ways---is the representation of the impurity as a (pseudo) spin-1/2 with the $xy$-coupling to the bosonic environment \cite{Whitsitt_Sachdev,kun2}. 
From now on, we will be assuming such a setup when referring to the bosonic counterpart of the critical superconductor with a solenoind. 
Here the unbiased spin-1/2 impurity corresponds to the half-integer-flux solenoid, while biasing the spin by local ``magnetic field" in the $z$-direction 
is equivalent to driving the solenoid away from the half-integer-flux regime.

As we noted above, microscopic vortex loops do not contribute to the magnetic flux of the impurity. The same is true for the microscopic worldline loops,
which do not contribute to the charge of the static impurity. Hence, the correspondence between the magnetic flux and the particle charge trapped by the impurity
is exact. This allows us to interpret the recent bosonic results \cite{kun1,Whitsitt_Sachdev,kun2} 
in terms of magnetic flux, arriving at the following picture. 
For the half-integer bare solenoid flux, there is no net screening effect: The total magnetic flux is exactly equal to that of the solenoid. The simplicity of this result is rather deceptive. It reflects merely the symmetry of the problem with respect to half-integer values of the solenoid flux, but not the role of the coupling between the solenoid and the critical environment.  In fact, there is a non-trivial critical coupling between the flux of the solenoid and the fluxes carried by the vortex lines. This coupling is responsible for the halon effect taking place on a slight departure of the solenoid flux from the half-integer value. No matter how small is the deviation
from the half-integer value of the flux, the critical environment generates and additional magnetic flux that completes the net flux to the nearest integer value (in the units of magnetic flux quantum).  The net magnetic flux is thus generically quantized even at the critical temperature.

{\it The magneto-halon effect. } On the approach to the half-integer value of the bare solenoid flux, the magnetic flux generated by the environment comes in the form of a large halo around the solenoid
The radius of the halo, $r_0$, diverges following the critical law \cite{kun1,Whitsitt_Sachdev,kun2} 
\begin{equation}
r_0 \, \propto \,   |\Delta \Phi |^{-\tilde{\nu}} , \qquad \tilde{\nu} = 2.33(5) .
\label{r_0}
\end{equation}
Here $\Delta \Phi$ is the small deviation of the bare solenoid flux from a half-integer value. 
When $r_0$ is much larger than the radius of the solenoid, the magnetic fluxes of the solenoid and the system are well separated. With the same 
(arbitrarily high at $r_0 \to \infty$) accuracy, the flux of the halo equals $\pm 1/2$. In this sense, one can speak of the critical fractionalization of the magnetic
flux. 

\begin{figure}[bht]
\includegraphics[width=0.8\linewidth]{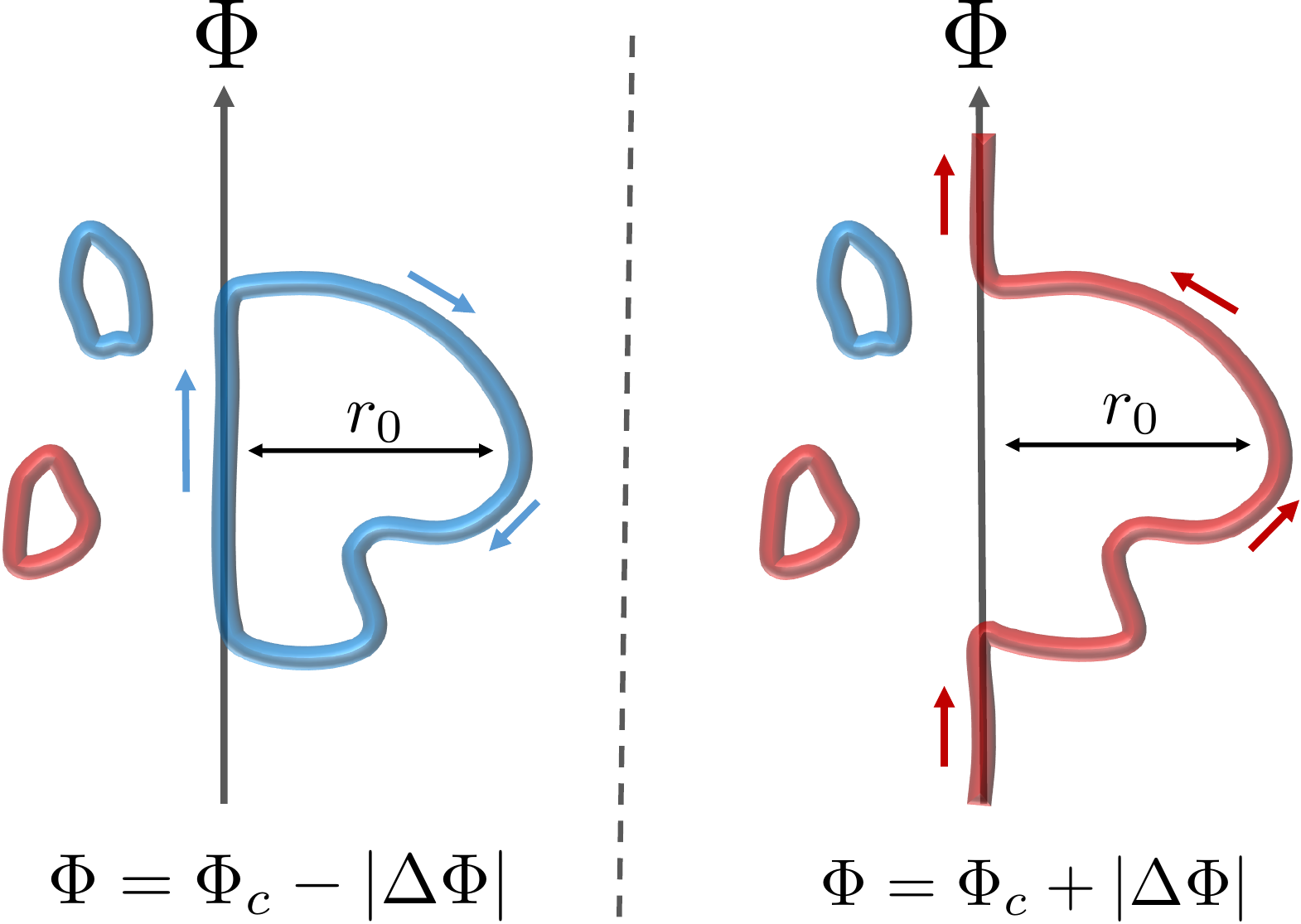}
\caption{The crucial role of large vortex loops in the magneto-halon effect in a critical superconductor. The black solid line represents a solenoid inserted into the system. When the bare solenoid flux $\Phi$ is close to half-integer number of flux quanta, $\Phi_c=(M+1/2)\, \Phi_0$ ($M=0$ in this plot), a few vortex loops with a large radius $\sim r_0$ [see Eq.~(\ref{r_0})] emerge, bringing on average and extra (plus or minus) one half of the flux quantum,  $\pm \Phi_0/2$.  The net effect is that one half of the flux quantum gets delocalized into a halo-shaped screening cloud with the critically divergent radius $r_0$. The blue and the red vortex loops, while carrying one flux quantum each, differ by the direction of the flux, as indicated by the arrows.}
\label{fig:halon}
\end{figure}

In Fig.~\ref{fig:halon}, we illustrate the crucial role of large vortex loops in the magneto-halon effect. Unlike their $(2+1)$-dimensional counterparts---bosonic worldlines, the 
vortices are real physical objects.  Therefore, the magneto-halon effect in 3D potentially allows a direct experimental insight into the origin of the halo.

{\it Flux-loop model with quasi-solenoid perturbation.}  In the case of the halon effect in the 2D U(1) quantum-critical environment, 
minimalistic lattice models---optimal, in particular, for numeric simulations---are the J-current models with a pseudo-spin-1/2 impurity \cite{Whitsitt_Sachdev,kun2}. 
While capturing all the universal properties of the magneto-halon effect, such models, however, do not posses one very characteristic 
feature of the U(1) gauge theory with a solenoid. Namely, that all the system's properties stay exactly the same if the bare magnetic flux of the solenoid is changed by an integer number of flux quanta. It would be thus very instructive to have a {\it quasi-solenoid} minimalistic model that, on one hand, would
demonstrate such a feature, while, on the other hand, would be as simple and convenient for simulations as a J-current model with pseudo-spin-1/2 impurity. 

We identify such a model, which turns out to be nothing but the textbook
``frozen lattice superconductor" \cite{Herbut} augmented with the  quasi-solenoid perturbation. We take a liberty of renaming the model, referring to it as the {\it flux-loop model}. This way we emphasize the following crucial property---especially relevant for our purposes---distinguishing the model from generic lattice gauge theories.  As we discussed above, the relationship between the quantization of vorticity and the quantization of the magnetic flux along a vortex loop is emergent, applying only to an infinitely large loop. 
In the flux-loop model, however, this property is hardwired at the microscopic level: The magnetic flux per plaquette is forced to be quantized 
by a rather special delta-functional form of the coupling term. 
Remarkably, this very specific physics property leads to substantial mathematical simplifications. With a generic lattice gauge theory, one arrives at the vortex-loop representation 
by a double duality transformation, going  first from the original variables to the (dual) representation of integer, divergence-free currents, and then performing the duality transformation from the currents to vortex loops. The unique simplicity of the flux-loop model allows one to directly reformulate the theory in terms of loops of quantized magnetic 
flux, without resorting to the two duality transformations and, correspondingly, without explicitly dealing with quantized vorticity. It is sufficient to perform a straightforward
integration over the vector potential in the partition function and then parameterize the result in terms of the integer numbers of quanta of magnetic fluxes per plaquette.

To see how the flux-loop model is special compared to a generic lattice gauge theory, it is instructive to briefly review a lattice version of the Ginzburg Landau theory in the extreme type-II limit, when the amplitude fluctuations of the order parameter are completely suppressed. The Hamiltonian and the partition function of the model read (see, e.g., texts~\cite{book,Herbut}):
\begin{equation}
\label{eq:H_LSM}
H_{\rm LSM}=\frac{1}{2e^2}\sum_{\bf{n}}(\nabla\times {\bf{A}_n})^2+\sum_{{\bf{n}}\alpha}f(\phi_{\bf{n}+\hat{\alpha}}-\phi_{\bf{n}}-A_{\bf{n} \alpha}),
\end{equation}
\begin{equation}
\label{eq:Z_LSM}
Z_{\rm LSM}=\int \! \mathscr{D}{\bf A} \mathscr{D}\phi \, e^{-H_{\rm LSM}[\phi, {\bf{A}}]} .
\end{equation}
Here $\phi_{\bf{n}}$ is the phase of the order parameter on the site $\bf{n}$ of the cubic lattice; $A_{\bf{n} \alpha}$ is the $\alpha$-component [$\alpha=(x,y,z)$] of the lattice vector potential, living on the bond $(\bf{n},\alpha)$. The lattice curl $\nabla\times {\bf{A}}$ defines the magnetic flux per plaquette. The only condition on the function $f(\phi)$---implied by the U(1) invariance of the theory---is the $2\pi$-periodicity. In the model (\ref{eq:H_LSM})--(\ref{eq:Z_LSM}),  the variables are scaled in such a way that both the temperature and the strength of coupling of the vector potential to the phase field are equal to unity, so that the square of the electric charge, $e^2$, remains the only control parameter of the theory.

Increasing $e^2$ drives the system from the (low-temperature) superconductor phase to the (high-temperature) normal phase, through the phase transition of the inverted 3D XY type. As long as the universal physics is concerned, the specific form of the function $f(\phi)$ is not important (for a detailed discussion, 
see, e.g., Ref.~\cite{book}). In the case of the flux-loop model (a.k.a. the frozen lattice superconductor \cite{Herbut}), one adopts the simplest choice 
\begin{equation}
\label{eq:simplest}
e^{- f(\phi)}=\sum_{M}\, \delta (\phi-2\pi M) ,
\end{equation}
with $M$ an integer.

The delta-functional structure of Eq.~(\ref{eq:simplest}) imposes the constraint $A_{\bf{n} \alpha} = \phi_{\bf{n}+\hat{\alpha}}-\phi_{\bf{n}} -2\pi M$, leading to
quantization of the magnetic field ${\bf B_n}=\nabla\times {\bf{A}_n}$ in integer multiples of $2\pi$. Adopting simple convention that the plaquette area equals unity allows
us to interpret the field ${\bf B_n}$ as the field of quantized magnetic flux per plaquette.

It is easy to see that  the set of all the (divergence-free) configurations of quantized field ${\bf B_n}$ exhausts all the physical degrees of freedom of the system. Indeed, for each lattice site, we have three independent delta-functions and four independent continuous variables (three components of the vector potential and the phase variable). This means that the integration in Eq.~(\ref{eq:Z_LSM}) removes all the delta-functions of Eq.~(\ref{eq:simplest}), still leaving one continuous variable per site. Upon removing the
delta-functions, the only terms left in the action are the squares of quantized magnetic fields/fluxes, $(\nabla\times {\bf{A}_n})^2 = {\bf B}_{\bf n}^2$. In view of their discrete
nature, these cannot depend on the continuous variables left upon the integration, bringing us to the conclusion that those continuous variables represent nothing but physically irrelevant gauge freedom.

Hence, the flux-loop model is exhaustively parameterized with the quantized, divergence-free magnetic(flux) field ${\bf B_n}$, and the result is the standard Villain model (with natural rescaling ${\bf B_n} \to 2\pi {\bf B_n}$, $e \to 2\pi e$, the quantum $2\pi$ gets absorbed into $e$ and the field values become integer)
\begin{equation}
\label{eq:V}
Z_{V}=\sum_{\{ {\bf B} \} }^{\nabla \cdot {\bf B}=0}  e^{-H_{V}[{\bf B}]},  \qquad H_{V}=\frac{1}{2 e^2} \sum_{\bf n } \, {\bf B_n}^2.
\end{equation}

Given that the very same model (\ref{eq:V}) can be derived from (\ref{eq:H_LSM})--(\ref{eq:simplest}) by the (double) duality transformation, in which case the field ${\bf B_n}$ would have the meaning of the quantized circulation of the phase gradient---thus representing a vortex, we conclusde that in the flux-loop model, the quanta of vorticity can also be viewed as the quanta of (quantized) magnetic flux per corresponding plaquette.

The above treatment also applies to the flux-loop model with external electric currents ${\bf J}_{\bf n}$ living on bonds:
\begin{equation}
\label{eq:solenoid}
H_{\rm LSM} \rightarrow H_{\rm LSM}-\frac{1}{e^2}\sum_{\bf n} \, {{\bf J}_{\bf n}\cdot {\bf A_ n}}.
\end{equation}
The resulting Villain model acquires an external magnetic field/flux ${\bf H}_{\bf n}$  living on plaquettes:
\begin{equation}
\label{eq:quasi_solenoid}
H_{V} \rightarrow \frac{1}{2 e^2} \sum_{\bf n} \, ( {\bf B}_{\bf n} - {\bf H}_{\bf n})^2 + {\rm const}.
\end{equation}
The field ${\bf H}_{\bf n}$ is found by solving $\nabla \times {\bf H}_{\bf n}={\bf J}_{\bf n}$. 
\begin{figure}[bht]
\includegraphics[width=0.5\linewidth]{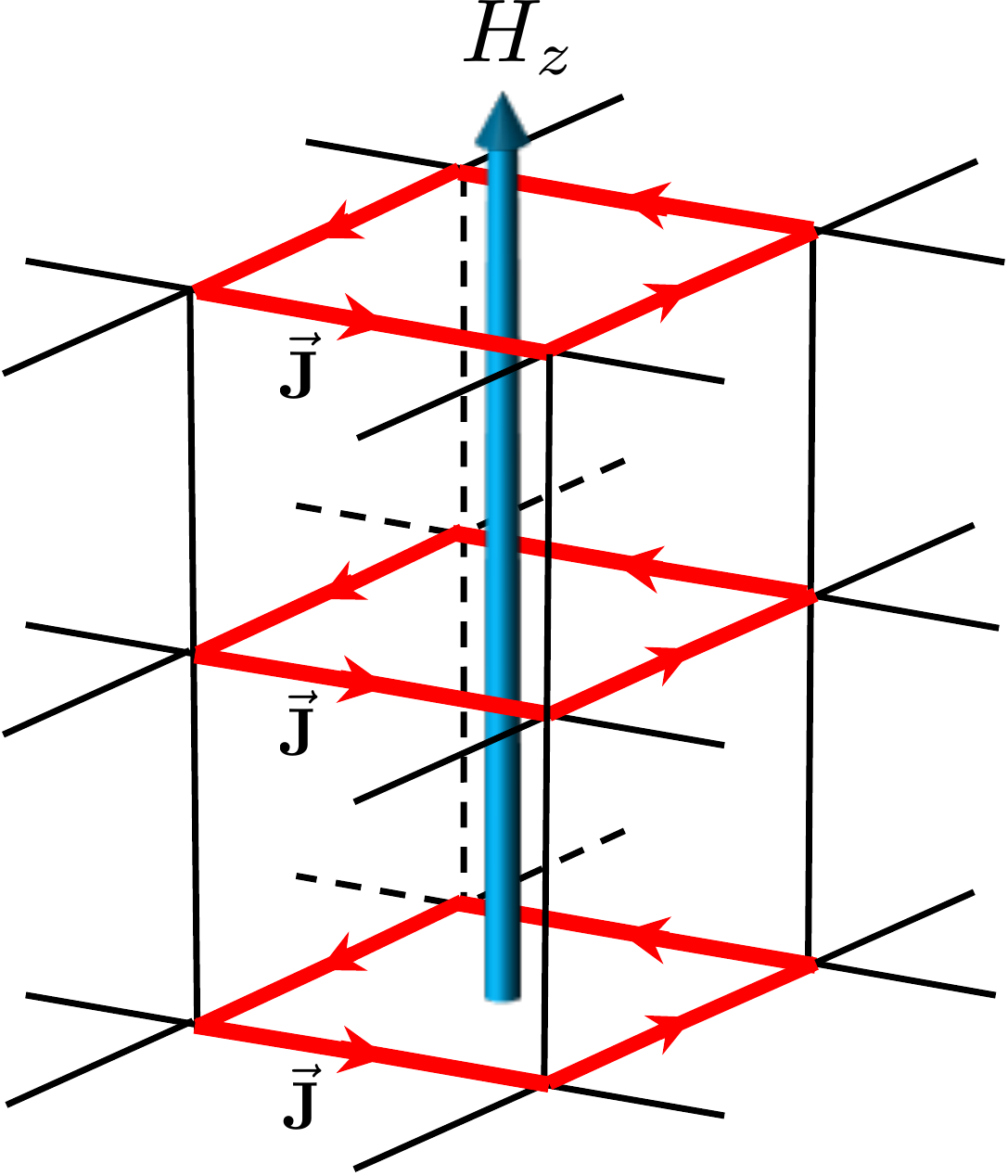}
\caption{The quasi-solenoid perturbation on the cubic lattice: Stacked current loops along a tube in the $z$-direction generate a magnetic field/flux $H_z$ within the tube.
In the flux-loop model, the quasi-solenoid perturbation possesses all the special symmetries of an ideal (infinitesimally thin) continuous-space solenoid.}
\label{fig:solenoid}
\end{figure}
With the setup of the quasi-solenoid perturbation, the external currents ${\bf J}_{\bf n}$ are stacked current loops along a tube in the $z$ direction, as shown in Fig.~\ref{fig:solenoid}. Corresponding field ${\bf H}_{\bf n}$ mimics the bare solenoid flux $\Phi$ of Eq.~(\ref{r_0}) and Fig.~\ref{fig:halon}.

 Let us see how our quasi-solenoid perturbation relates to an ideal solenoid. In the absence of environment, there is no difference: the magnetic field is zero everywhere except for the special line of bonds with biasing field. With the environment, the difference comes from the fact that the special line of bonds is a part of the system, while in the case of an ideal (infinitesimally thin) solenoid, the perturbation is nothing but a gauge phase. Despite the fact that our perturbation does not reduce to a pure gauge phase, corresponding qualitative features are essentially the same: Equation~(\ref{eq:quasi_solenoid}) implies that the systems's properties, as a function of the strength of the solenoid biasing field $H_{{\bf n}_0 z} $  (here ${\bf n}_0$ stands for the solenoid sites), are exactly periodic, except for the shift of the magnetic flux within the solenoid by an integer number of quanta. In particular, there are trivial points $H_{{\bf n}_0  z} = 2\pi M $---analogs of the trivial gauge phase  $2\pi  M$---where the properties of the system are exactly the same as in the absence of bias. Finally, there are special (and microscopically equivalent to each other) points, $H_{{\bf n}_0 z} = 2\pi (M+1/2)$,---analogs of the gauge phase $2\pi (M+1/2)$---where the halon boundary phase transitions take place.

{\it Conclusions.}  The problem of magnetic flux quantization can be formulated without regard to the state of the system (superconducting, normal, or critical): 
It is always legitimate to consider a perturbation created by inserted thin solenoid and ask the question of whether the net magnetic flux---the bare flux of the solenoid 
plus the flux induced in the environment---is quantized or not. With such a formulation, 
we addressed the finite-temperature critical state of a 3D superconductor. By the well-known vortex-worldline mapping, the problem reduces to the recently solved problem of quantization of the charge of a static impurity  in the 2D U(1) quantum-critical environment. 

This way we see that at the critical temperature, the net magnetic flux is generically quantized,
a dramatic difference with the standard superconducting case taking place on the approach to the critical (half-integer) values of the bare magnetic flux of the solenoid. 
The half-integer values (in the units of magnetic flux quantum), $M+1/2$, of the bare flux correspond to the boundary phase transitions between two discrete values ($M$ and $M+1$) of the net magnetic flux. On the approach to the $(M+1/2)$-point, the magneto-halon effect develops. The integer net magnetic flux gets fractionalized into two parts: a microscopic core with the flux $M+1/2$---the bare flux of the solenoid---and a critically large [see Eq.~(\ref{r_0})] halo carrying a complementary charge of $\pm 1/2$. 
Right at the $(M+1/2)$-point, the halo disappears (becoming infinitely large) and the net magnetic flux equals $(M+1/2)$.

As illustrated in Fig.~\ref{fig:halon}, the magneto-halon effect is due to a very special behavior of large vortex loops in. Unlike their $(2+1)$-dimensional counterparts---bosonic worldlines, the vortices are real physical objects, potentially allowing a direct experimental insight into the origin of the effect.

Finally, we identified the flux-loop model (a.k.a. frozen lattice superconductor) with a quasi-solenoid perturbation as the model of choice for the halon and the magneto-halon effects. Unlike a generic model of the same universality class, the flux-loop model has a very special microscopic property---the quantization of magnetic flux per plaquette.
This feature leads to the most straightforward parameterization of the model in terms of the quantized magnetic fluxes. In this parameterization, the model acquires the form of the standard Villain model. The quasi-solenoid perturbation in the flux-loop model possesses all the special symmetries of an ideal (infinitesimally thin) continuous-space solenoid. 
These symmetries allow to immediately identify the points of the (magneto-)halon boundary phase transitions, which is particularly valuable for numeric studies.

K.C. acknowledges helpful discussions with Yashar Komijani and Po-Yao Chang. This work was supported by the National Science Foundation under the grant DMR-1720465 and the MURI Program ``New Quantum Phases of Matter" from AFOSR.


\begin{thebibliography}{99}

\bibitem{book} B. Svistunov, E. Babaev, and N. Prokof'ev,  {\it Superfluid States of Matter}, Taylor \& Francis,  2015.

\bibitem{Halperin1}
B. I. Halperin, T. C. Lubensky, and S. K. Ma,
{Phys. Rev. Lett.} {\bf 32}, 292 (1974).

\bibitem{Halperin2}
C. Dasgupta and B. I. Halperin,
 {Phys. Rev. Lett.}  {\bf 47}, 1556 (1981).

\bibitem{kun1} Y. Huang, K. Chen, Y. Deng, and B. Svistunov, Phys. Rev. B {\bf 94}, 220502 (2016).

\bibitem{Whitsitt_Sachdev} S. Whitsitt and S. Sachdev, Phys. Rev. A {\bf 96}, 053620 (2017).

\bibitem{kun2} K. Chen, Y. Huang, Y. Deng, and B. Svistunov, arXiv:1807.02168.

\bibitem{Herbut} I. Herbut, {\it A Modern Approach to Critical Phenomena}, Cambridge University Press, 2007.




\end{thebibliography}
\end{document}